\documentstyle[prl,aps]{revtex}
\begin{document}
\title{Size Gap for Zero Temperature Black Holes in Semiclassical Gravity}
\author{Paul R. Anderson\cite{PRA} and Courtney D. Mull\cite{CDM}}
\address{Department of Physics\\Wake Forest University\\Winston-Salem, NC  27109}
\maketitle
\begin{abstract}
We show that a gap exists in the allowed sizes of all zero temperature static spherically
symmetric black holes in semiclassical gravity when only conformally invariant fields
are present.  The result holds for both charged and uncharged black holes.  By size we
mean the proper area of the event horizon.  The range of sizes that do not occur depends
on the numbers and types of quantized fields that are present.  We also derive some general
properties that zero and nonzero temperature black holes have in all classical and semiclassical metric theories of gravity.
\end{abstract}

One of the unanswered questions in semiclassical gravity is how quantized fields alter the spacetime geometry near the event horizon of a black hole.  This is important because the thermodynamic properties of a black hole are determined by the geometry at the event horizon.  Some work has been done to answer this question using linearized semiclassical gravity and either analytical approximations or numerical computations of the stress-energy tensor in Schwarzschild spacetime\cite{York,HK,HKY,AHWY,Winnipeg}.  Further progress has been hampered by problems with the analytical approximations near the event horizons of other black holes\cite{FZ,AHS,AHL} and by the difficulty involved in numerically computing the stress-energy tensor for quantized fields in black hole spacetimes.  

Given these difficulties and the importance which quantum effects may have in black hole spacetimes, it is useful to see if it is possible to deduce anything about the general
properties of static spherically symmetric black holes without solving the full 
nonlinear set of semiclassical equations.  We show that significant restrictions on the
spacetime geometry near the event horizon of a black hole can be obtained by just requiring 
that no scalar curvature singularities exist at the horizon.  These results apply to all
classical and semiclassical metric theories of gravity.  By requiring that the spacetime
be a solution to the trace of the semiclassical backreaction equations we show that
there exists a range of sizes for which no zero temperature black holes exist, so long as
only conformally invariant fields are present.  This result also applies to some possible
nonzero temperature black holes.  By size we mean the proper area of the event horizon.
The range of excluded sizes depends on the number and types of quantized fields present.

Some previous work has been done to determine the general properties black holes
must have in certain situations.  Mayo and Bekenstein \cite{MB} considered
static spherically symmetric black hole solutions to Einstein's equations for various
types of matter fields.  They found that black hole solutions exist if the stress-energy tensor is finite on the horizon and one component satisfies a certain inequality.  The
geometry near the horizon is of the same form as the Schwarzschild geometry except in the
limit of an extreme black hole where the inequality becomes an equality.  In this latter case they put constraints on the form of the geometry near 
the horizon.  Recently Zaslavskii \cite{Z} has used a power series expansion of the metric to determine the
general form it takes near the horizons of near extreme and extreme charged
black holes in a cavity when the grand canonical ensemble is utilized.

There has also been some previous evidence of excluded sizes for black holes when quantum
effects are taken into account.  Peleg, Bose, and Parker\cite{PBP} have found the existence 
of a mass gap in the formation of black holes in two-dimensional dilaton theories of gravity.  
Brady and Ottewill\cite{BO} have recently found evidence that quantum effects can cause a 
mass gap in the formation of black holes in four dimensional spherically symmetric gravitational 
collapse.

We begin by examining the properties that result from requiring the spacetime curvature to
be finite at the event horizon.  These properties apply to black hole geometries in
any classical or semiclassical metric theory of gravity.  The metric for a static spherically symmetric spacetime can be written in the general form\cite{units}
\begin{equation}
d s^2 = - f(r) dt^2 + \frac{1}{k(r)} dr^2 + r^2 d \Omega^2 \;\;\;.
\end{equation}
If the spacetime has an event horizon then $f$ vanishes on that horizon and
the surface gravity is given by the formula 
\begin{equation}
\kappa = \frac{v}{2} \left(f k \right)^{1/2} \;\;.
\end{equation}
The unique non vanishing components of the Riemann curvature tensor in an orthonormal frame are
\begin{mathletters}
\begin{eqnarray}
R_{\hat{t} \hat{r} \hat{t} \hat{r}} &=&  \frac{v' k}{2} + \frac{v k'}{4} + \frac{v^2 k}{4} \\
R_{\hat{t} \hat{\theta} \hat{t} \hat{\theta}} &=&  \frac{v k}{2 r} \\
R_{\hat{r} \hat{\theta} \hat{r} \hat{\theta}} &=& - \frac{k'}{2 r} \\
R_{\hat{\theta} \hat{\phi} \hat{\theta} \hat{\phi}} &=& \frac{1 - k}{r^2}\;\;,
\end{eqnarray}
\end{mathletters}%
where $v \equiv f'/f$ and primes denote derivatives with respect to $r$.
The Kretschmann
scalar for the metric (1) is 
\begin{equation}
  R_{\alpha\beta\gamma\delta} R^{\alpha\beta\gamma\delta} = 
  4 (R_{\hat{t} \hat{r} \hat{t} \hat{r}})^2 + 
  8 (R_{\hat{t} \hat{\theta} \hat{t} \hat{\theta}})^2  +
  8 (R_{\hat{r} \hat{\theta} \hat{r} \hat{\theta}})^2 +
  4 (R_{\hat{\theta} \hat{\phi} \hat{\theta} \hat{\phi}})^2
\end{equation}
Thus even though the coordinate system in (1) is singular on the event horizon, it
is necessary that right hand sides of Eqs.(3a) - (3d) each be separately finite there
to avoid a scalar curvature singularity. 

To determine properties of $f$ and $k$ that arise from the requirement that the
curvature be finite first note that, since $f = 0$
on the horizon, $v = \infty$ there.  Thus from Eq.(3b) one can
infer that $k = 0$ at the horizon, while from Eq.(3c) it is clear that 
$k'$ is finite there.  The most significant constraints come from Eq.(3a).  If
the horizon is at $r = r_0$, then this equation can be written 
\begin{equation}
\frac{v' k}{2} + \frac{v k'}{4} + \frac{v^2 k}{4} = A(r)
\end{equation}
with $A(r)$ the function obtained by substituting specific functions $k$ and $f$
into Eq.(3a).  Solving this equation for $k'$ and noting that $A$ is finite at
the horizon, one finds that if $k' = c_2$ at the horizon, with $c_2$ some positive constant,
then $k = c_2(r-r_0)$ and the equation becomes to leading order
\begin{equation}
 1 = - (v + 2 v'/v)(r-r_0)\;\;.
\end{equation}
The only solutions to this equation for which $f$ vanishes are $f = c_1 (r-r_0)$, with
$c_1$ a positive constant.  Thus either 
\begin{equation}
   f = c_1 (r-r_0) \;,\;\;\; k = c_2 (r-r_0) \;\;,
\end{equation}
near the horizon, or $k' = 0$ at the horizon.  One can also formally integrate Eq.(5) to
obtain
\begin{equation}
  k = \frac{B_0}{v^2 f} + \frac{4}{v^2 f} {\int_{r_0}}^r f'(r_1) A(r_1) d r_1 \;\;.
\end{equation}
Here $B_0$ is a nonnegative constant.  Multiplying by $v^2 f$, noting that the second
term on the right is then zero at the horizon, and comparing with Eq.(2) gives
$B_0 = 4 \kappa^2$.  Multiplying Eq.(8) by $v^2$ shows that for zero temperature black 
holes $k v^2$ is finite and thus that $k v = 0$ at the horizon.  
For nonzero temperature black holes $kv = 4 \kappa^2/f'$ at the horizon.  Thus since
$kv$ must be finite, $f'$ cannot vanish at the horizon.  If $f'$ diverges there
then $kv = 0$.  If $f'$ approaches a constant then it is easy to see that
Eq.(7) describes the behavior of $f$ and $k$ near the horizon. Thus for all black holes,
either the geometry near the horizon is of the form (7) or at the horizon $k' = kv = 0$.

For nonzero temperature black holes the proper
distance to the event horizon along a radial geodesic  
is finite.  To see this note that this distance is 
\begin{equation}
 {\int_{r_0}}^r d r_1/(k(r_1))^{1/2} \;\;.
\end{equation}
Solving Eq.(2) for $k$, substituting into (9) and integrating one finds that 
the integral is always finite.  This result also allows one to make the change of variable 
$d l = dr/k^{1/2}$.  With this change of variable Eq.(5) becomes
\begin{equation}
\frac{1}{2 f} \frac{d^2 f}{d l^2} - \frac{1}{4 f^2} \left(\frac{d f}{d l}\right)^2 
  = A(l) \;\;.
\end{equation}
Multiplying by $f/(df/dl)$ and integrating one finds that, 
since $A$ is finite at the horizon, the only solutions for which
$f$ vanishes are of the form
\begin{equation}
  f = c_1 (l - l_0)^2 \;\;
\end{equation}
near the horizon.  Here $l = l_0$ at the horizon and $c_1$ is a positive constant.

To find the proper distance to the horizon for zero temperature black holes note 
that near the horizon $k \le c_2/v^2$ for some $c_2>0$ since $k v^2$ is finite at the horizon.
Substituting this inequality into Eq.(5) and integrating one finds that the proper distance is always infinite for zero temperature black holes.  

It is now possible to show that for all zero temperature black holes $\Box R \ne a_0$ 
at the horizon for any nonzero constant $a_0$.  First note that one can always write the
equation $\Box R = a(r)$ with $a(r)$ the function which results from computing $\Box R$
for a particular metric.  We will assume that $a(r_0) = a_0 \ne 0$ at 
the horizon and then show that this is not possible without there being a curvature
singularity there.  For the metric (1) the equation $\Box R = a(r)$ can be formally 
integrated with the result that to leading order near the horizon
\begin{equation}
  R = b_1 + b_2 {\int_{r_0}}^r d r_1 (k(r_1) f(r_1))^{-1/2} +  
           a_0 {\int_{r_0}}^r d r_1 (k(r_1) f(r_1))^{-1/2}
          {\int_{r_0}}^{r_1} d r_2 (f(r_2)/k(r_2))^{1/2}\;\;.
\end{equation}
Here $b_1$ and $b_2$ are arbitrary constants.  For all zero temperature black holes (9) 
diverges and therefore $R$ diverges unless $b_2 = 0$.   Since $k v^2$ is finite on the 
horizon for zero temperature black holes, $k^{-1/2} \ge c_2 v$ for some $c_2 > 0$.  
Substituting this inequality into the last integral on the right in Eq.(12) gives
\begin{equation}
  |R - b_1| \ge  2 |a_0| c_2 {\int_{r_0}}^r d r_1/(k(r_1))^{1/2} \;\;.
\end{equation}
Since the integral diverges, $R$ must diverge on the horizon which means there is a
curvature singularity there.

For nonzero temperature black holes one can make the variable transformation $dl = dr/k^{1/2}$
in Eq.(12).  Assuming $a = a_0$ at the horizon and substituting Eq.(11) into Eq.(12), one
finds to leading order the following equation for $R$,
\begin{equation}
R = -\frac{4}{r_0} \frac{d^2 r}{d l^2} - \frac{4}{r_0 (l-l_0)} \frac{d r}{d l} + 
     \frac{2}{r_0^2} = b_1 + \frac{a_0}{4}(l - l_0)^2 \;\;.
\end{equation} 
Note that $b_2$ must still be zero for $R$ to be finite.  This equation can be solved 
to give the
only two forms the metric can have near the horizon that could actually result in 
$\Box R = a_0$ there.  They are either Eq.(7) or
\begin{equation}
f = c_1 (r - r_0)^{1/2} \;\;,\;\;\; k = c_2 (r - r_0)^{3/2} 
\end{equation}
with $c_1$ and $c_2$ positive constants.  All other possible black hole metrics
result in $\Box R$ being either divergent or zero at the horizon.

Semiclassical gravity can be used to place further constraints on the geometry near the event horizon of a static spherically symmetric black hole in the case that only conformally invariant free quantized fields are present.  The semiclassical backreaction equations can be written in the general form
\begin{equation}
G_{\mu\nu} + u U_{\mu\nu} + w W_{\mu\nu} = 8 \pi (T_{\mu\nu})_{cl} + 8 \pi <T_{\mu\nu}>
\end{equation}
Here $U_{\mu\nu}$ and $W_{\mu\nu}$ are tensors which result from the variation of an $R^2$ term and a $C_{\alpha\beta\gamma\delta} C^{\alpha\beta\gamma\delta}$ term in the gravitational Lagrangian, with $C_{\alpha\beta\gamma\delta}$ the Weyl tensor.  Their coefficients, $u$ and $w$, are arbitrary and must in principle be determined by experiment or observation.  $(T_{\mu\nu})_{cl}$ is the stress-energy tensor for any classical fields.  We shall only be concerned here with the classical electromagnetic field\cite{note1}.  The trace of  $W_{\mu\nu}$ is identically zero as is the trace of the stress-energy tensor for the classical electromagnetic field.  The trace of $U_{\mu\nu}$ is equal to $-6 \Box R$\cite{BD}.  For conformally invariant fields the trace of $<T_{\mu\nu}>$ is equal to the trace anomaly\cite{BD}.  Thus the trace equation is
\begin{equation}
- R - 6 u \Box R = 8 \pi [\alpha \Box R + \beta (R_{\alpha\beta} R^{\alpha \beta} - \frac{1}{3} R^2) + \gamma   
       C_{\alpha\beta\gamma\delta} C^{\alpha\beta\gamma\delta}]
\end{equation}
with
\begin{mathletters}
\begin{eqnarray}
\alpha &=& [N(0) + 6 N(1/2) - 18 N(1)]/2880 \pi^2 \\
\beta &=& [N(0) + 11 N(1/2) + 62 N(1)]/2880 \pi^2 \\
\gamma &=& [N(0) + \frac{7}{2} N(1/2) - 13 N(1)]/2880 \pi^2 \;\;.
\end{eqnarray}
\end{mathletters}%
Here $N(0)$, $N(1/2)$, and $N(1)$ are the number of scalar, four component spin $1/2$, and vector fields respectively. 

Since all of the other terms in Eq.(17) are finite at the horizon, $\Box R$ must be finite
as well.  Then for all zero temperature black holes and for all nonzero temperature ones with
metrics near the horizon other than those given by Eqs.(7) and (15), $k' = kv = \Box R = 0$ at
the horizon.  
The components of the Riemann tensor there are then $R_{\hat{t} \hat{r} \hat{t} \hat{r}} = 
A(r_0) = A_0$, $R_{\hat{t} \hat{\theta} \hat{t} \hat{\theta}} = 0$,
$R_{\hat{r} \hat{\theta} \hat{r} \hat{\theta}} = 0$, and $R_{\hat{\theta} \hat{\phi} \hat{\theta} \hat{\phi}} = 1/r_0^2$.  The terms in Eq.(17) can be computed at the horizon 
from these components.  After some algebra we find the following equation for $A_0$.
\begin{equation}
A_0^2 \left(\frac{16 \pi}{3}(\beta+ 2 \gamma)\right) + A_0 \left(-2 + \frac{64 \pi}{3 r_0^2}
    (\beta-\gamma)\right) + \left(\frac{2}{r_0^2} + \frac{16 \pi}{3 r_0^4} (\beta+2 \gamma)
    \right) = 0
\end{equation}
Since $\beta+ 2 \gamma > 0$ for all fields there is no solution
to this equation if $A_0 = 0$.  For zero temperature black holes this result along with
Eq.(8) implies that near the horizon
\begin{equation}
  k = 4 A_0/v^2 \;\;.
\end{equation}
For $A_0 \ne 0$, Eq.(19) can be solved with the result that
\begin{equation}
A_0 = \frac{1}{16 \pi (\beta + 2 \gamma) {r_0}^2} \, [3 {r_0}^2 - 32 \pi (\beta - \gamma) \pm (768 \pi^2 \beta^2 - 3072 \pi^2 \beta \gamma - 288 \pi \beta {r_0}^2 + 9 {r_0}^4 )^{1/2}] \;.
\end{equation}
$A_0$ is not real (and therefore there can be no solutions to the semiclassical
backreaction equations with $k' = kv = \Box R = 0$ at the horizon) if  
\begin{eqnarray}
r_{-} &<& r_0 <r_{+} \nonumber \\
 r_{\pm} &=& 4 (\pi \beta)^{1/2} \left[1 \pm \left(\frac{2}{3 \beta} \right)^{1/2} (\beta + 2 \gamma)^{1/2} \right]^{1/2} \;\;\;.
\end{eqnarray}
Thus only nonzero temperature black holes with the
behaviors (7) or (15) near the horizon can exist with sizes in this range.  
All other black holes
including all zero temperature black holes cannot have sizes in this range.
For all allowed values of $\beta$ and $\gamma$\, $r_{+}$ is real.  If 
$\beta \le 4 \gamma$ then $r_{-}$ is imaginary or zero and solutions only occur for  
$r_0 \ge r_{+}$.  If $\beta > 4 \gamma$ then solutions also occur for $0 < r_0 \le r_{-}$.

\acknowledgments

P.\ R.\ A.\ would like to thank E. Carlson for helpful comments.  This work was supported in part by Grant No. PHY95-12686 from the National Science Foundation.


\begin{references}
\bibitem{note1} In fact the results will be valid for any classical stress-energy tensor whose trace vanishes.
\end{references}
\end{document}